# Local Diagnostics of the Density of Neutral Atoms in Plasma by Stimulated Rotation of Polarization Plane

Krzysztof Dzierżęga[1], Michal Hnatic[2,3,4],

Koryun B. Oganesyan[5,*], Peter Kopcansky[2]

[1] Marian Smoluchowski Institute of Physics, Jagiellonian University, Kraków, Poland

[2] Institute of Experimental Physics SAS, Kosice, Slovakia

[3] Joint Institute for Nuclear Research, Dubna, Russia
[4] Faculty of Sciences, P. J. Safarik University, Kosice, Slovakia

[5] A. Alikhanyan National Science Lab, Yerevan Physics Institute, Yerevan, Armenia

[*] bsk@yerphi.am

The polarization plane stimulated rotation angle of probe signal in the intense laser field in plasma is calculated. The estimates of the residual gas local density in a cesium plasma based on the effects of Faraday, Cotton-Mouton and stimulated rotation of probe signal in the intense laser field have been found. A brief theory of resonant change in the plane of polarization of a probe signal under the action of an intense pulse is presented. It is shown, that the rotation in the medium has a complex structure.

## 1. Introduction

One of the main tasks set before plasma researchers is the problem of using thermonuclear energy [1]. At the first stages of development, the main attention was paid to the determination of such plasma parameters as the concentration of charged particles and (electrons and ions ), their temperatures and , the degree of ionization, etc. Now more and more attention of researchers is attracted by the problem of determining the residual density of the neutral component of the plasma and the magnitude of the magnetic field. Of particular importance is the determination of these parameters in tokamaks [2].

Plasma diagnostic methods can be classified according to different criteria. For example, depending on whether the sensitive sensor is in contact with the medium under study or not, it is customary to call measurement methods contact or non-contact. In the modern plasma experiment, there is a clear desire for non-contact diagnostics of plasma. In this case, both passive diagnostics are developed - the electromagnetic field or corpuscular radiation emitted by the object under study is analyzed, and active diagnostics - the probing beam of radiation or particles scattered by the plasma is analyzed [3-7].

Under normal conditions, the gaseous medium is an optically isotropic medium. Optical anisotropy in a gas can arise when the gas is placed in a magnetic field (Faraday and Cotton-Moutton effects), under the action of intense laser radiation of elliptical or circular polarization. Let us dwell on the last phenomenon.

Intense laser radiation passing through a gaseous medium. changes as the absorption coefficient. and the refractive index of the medium. If a wave has a degree of circular polarization other than zero, then the changes in the absorption coefficient and refractive index caused by it are different for different circular polarizations of light. The first circumstance leads to the induced circular dichroism of the medium. the second is to induced birefringence. If a test linearly polarized signal is passed through such a medium, then its polarization will change at the exit from the medium; due to dichroism, plane-polarized light will turn into elliptically polarized, due to birefringence, a rotation of the plane occurs. The above phenomena in a gas have a pronounced resonant character and become actually observable under the condition that the frequency of laser radiation is close to the frequency of the atomic transition [8].

The rotation of the plane of polarization of a weak signal under the action of a circularly polarized signal (under the conditions of one and two-photon resonances) was observed for the first time by the group of V.M. Arutyunyan [9]. Later, Liao and Berkland also observed two-photon resonant rotation of the plane of polarization in sodium vapor [10-13]. The change in the polarization of a weak signal in the field of a strong one formed the basis of super-resolution polarization spectroscopy, and the polarization method of investigation made it possible to improve the measurement accuracy by 2–3 orders of magnitude.

Theoretical study of the effect of resonant rotation of the polarization plane is the subject of a number of papers [8, 13-15 and references therein].

## 2. Resonant Change in the Plane of Polarization of a Probe Signal under the Action of an Intense Pulse

Let us consider the passage of elliptically polarized radiation with the electric vector

$$\vec{E}' = \vec{E}(z,t)e^{-i\omega t} + \vec{E}^*(z,t)e^{i\omega t} \qquad (1)$$

through a resonant medium, consisting of identical two-level atoms. In an isolated state, the energy levels of atoms are degenerate once in terms $2J+1$ of the projection M of the moment of quantity. In the field of polarized radiation, the degeneracy of atomic levels is lifted. And in reality the problem is reduced to the interaction of radiation with a set of multilevel atoms. For simplicity, we consider an atomic transition $J_1 = 1/2 \to J_2 = 1/2$.

The energy levels of atoms without a field are doubly degenerate in M (Fig. 1); in the field of the wave, the degeneracy is lifted .

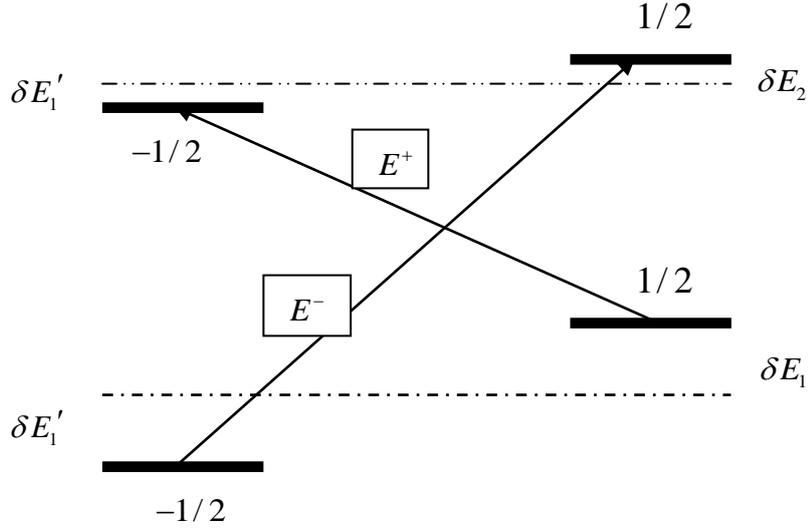

Fig.1. Resonance splitting and shift of energy levels of Cesium atoms (resonance transition $6S_{1/2} \to 6P_{1/2}$). The dotted line indicates the position of the levels in the absence of a electromagnetic field. Energy shifts $\delta E_1 = -\delta E_1' = \mu_0 H$, $\delta E_2 = -\delta E_2' = \dfrac{\mu_0 H}{3}$

The self-consistent system of equations for a resonant medium in field (1) has the form

$$\frac{\partial \rho_\pm}{\partial t} + v \frac{\partial \rho_\pm}{\partial z} - i\varepsilon \rho_\pm + \frac{\gamma}{2} \rho_\pm = -\frac{id}{\sqrt{G}\hbar} E_\pm \Delta_\pm,$$

$$\frac{\partial \Delta_\pm}{\partial t} + v \frac{\partial \Delta_\pm}{\partial z} + \frac{1}{\tau}(\Delta + 1/2) = \frac{2id}{\sqrt{G}\hbar} E_\pm \rho_\pm^* - \frac{2id^*}{\sqrt{G}\hbar}(E_\pm)^* \rho_\pm, \qquad (2)$$

$$\left(\nabla^2 - \frac{1}{c^2} \frac{\partial^2}{\partial t^2}\right) E_\pm = \frac{4\pi}{c^2} \frac{\partial^2}{\partial t^2} \left(\frac{2Nd^*}{\sqrt{G}} \int dv W(v) \rho_\pm \right),$$

where $\rho_\pm$, $\Delta_\pm$ are the transition current and overpopulation respectively of waves created by circular components $E_\pm = E_x \pm i E_y$, $W(v) = \dfrac{2\omega}{c\sqrt{\pi}\gamma_D} e^{-\frac{mv^2}{2kT}}$ is Doppler velocity distribution of atoms. $\gamma_D = \dfrac{2\omega}{c}\sqrt{\dfrac{2kT}{m}}$ is non-uniform Doppler width., $\gamma$ is the uniform width, $\tau$ - the lifetime of an atom, d is the reduced transition matrix element, $\varepsilon = \omega - \omega_0$ resonance detuning, N is the density of atoms.

Suppose, that the radiation field (1) consists of a strong wave of frequency $\omega$, propagating along the z axis and a weak wave with a frequency $\omega'$ propagating in the opposite direction

$$\vec{E}(z,t) = \vec{E}_s(z) e^{i\frac{\omega}{c}z} + \vec{E}_w(z) e^{-i\frac{\omega}{c}z - i(\omega'-\omega)t} \qquad (3)$$

Linearizing the system of equations (2) with respect to the weak field, in the stationary case, without taking into account coherent effects, we find

$$\frac{\partial E_s^{\pm}}{\partial t} = -i\frac{2\pi N\omega|d|^2}{3c\hbar} E_s^{\pm} \int dv \frac{W(v)(-\Delta_{\pm})}{(\varepsilon - \frac{\omega v}{c} + i\gamma/2)},$$

$$\frac{\partial E_w^{\pm}}{\partial t} = i\frac{2\pi N\omega|d|^2}{3c\hbar} E_w^{\pm} \int dv \frac{W(v)(-\Delta_{\pm})}{(\varepsilon' + \frac{\omega v}{c} + i\gamma/2)} \quad (4)$$

where $\varepsilon = \omega - \omega_0$ detuning of an intense wave, $\varepsilon' = \omega' - \omega_0$ detuning of a weak wave, $\Delta_{\pm} = -\frac{1}{2}\frac{(\varepsilon - \frac{\omega v}{c})^2 + \gamma^2/4}{(\varepsilon - \frac{\omega v}{c})^2 + \Gamma_{\pm}^2/4}$ is the overpopulation created by an intense wave, $\Gamma_{\pm}(z) = \gamma\sqrt{1 + G_{\pm}(z)}$ is a uniform width taking into account non-linear effects, $G_{\pm}(z) = \frac{4}{3\gamma^2}\gamma\tau\left|\frac{dE_s^{\pm}(z)}{\hbar}\right|^2$ - non-linear intensity parameter.

From the obtained system of equations (4) it is clear, that when passing through a resonant medium, the various circular components of the wave change in different ways, which leads to a change in the polarization of the radiation. If the strong wave is circularly polarized at the input (left circle), and the weak one is linear along the x-axis, then a y polarization component of the weak wave also appears at the output.

In the approximation of perturbation theory, the power $P_{wx}(z)$ of y polarization component of a weak wave at the output is determined by the formula

$$P_{wy}(z) = \frac{P_{wx}(0)}{4}\frac{G_+^2}{64}\frac{\gamma^2(2u\varepsilon + v\gamma)^2}{u^2\varepsilon^2(4\varepsilon^2 + \gamma^2)} e^{-\kappa_0 uz}\left(1 - e^{-\kappa_0 uz}\right)^2, \quad (5)$$

where $P_{wx}(0)$ is the power of x component at the input, z is the length of the resonant medium, $\kappa_0 = \frac{4\pi^{3/2}N|d|^2\omega}{3\hbar c\gamma_D}$, $u(\alpha,\beta) = \frac{1}{\pi}\int_{-\infty}^{+\infty}\frac{dt\beta e^{-t^2}}{(\alpha-t)^2 + \beta^2}$ and $v(\alpha,\beta) = \frac{1}{\pi}\int_{-\infty}^{+\infty}\frac{dt(\alpha-t)e^{-t^2}}{(\alpha-t)^2 + \beta^2}$ are probability integrals taken at the points $\alpha = \frac{2\varepsilon}{\gamma_D}$ and $\alpha = \frac{\gamma}{\gamma_D}$. We take $\varepsilon = \varepsilon'$ for the simplicity.

The probability integral (or Voigt function) is nothing more than a convolution of the Gaussian and Lorentzian line profiles. Physically, the probability integral u is related to the absorption of the wave, integral v - with phase or refractive index.

It can be seen from formula (5), that the linear polarization of a wave in a medium changes as a result of circular dichroism. So, it is due to the circular dispersion of the refractive index.

Changes in the polarization of the probe signal depend both on the shape of the spectral line (parameter $\beta$). So, it depends on the value of resonance detuning (parameter $\alpha$). Consider some possible rotation cases.

a) Doppler broadening prevails over homogeneous ($\gamma_D \gg \gamma$). For resonance detunings small compared to the Doppler width $\varepsilon \ll \gamma_D$, we have

$$P_{wy}(z) = \frac{P_{wx}(0)}{4} \frac{G_+^2}{16} \frac{\gamma^2}{4\varepsilon^2 + \gamma^2} e^{-\kappa_0 u z} \left(1 - e^{-\kappa_0 u z}\right)^2, \qquad (6)$$

Only atoms with detunings on the order of a uniform width contribute to the polarization change, which leads to the detection of a narrow homogeneous resonance on a wide Doppler contour. With increasing detuning, along with the central peak $\varepsilon \sim \gamma$, two side peaks appear $\varepsilon \sim \pm\gamma_D/2$, which is associated with an increase in the gyrotropy of the medium (Fig. 2.)

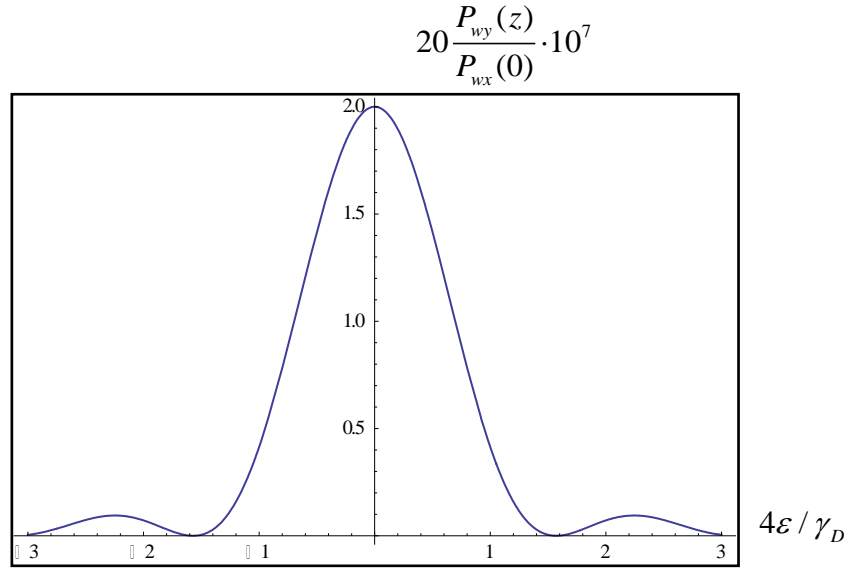

Fig.2. Spectral dependence of an elliptically weak wave at the output ($z = 20$ cm) at the value of the parameter $\beta = \gamma/\gamma_D \sim 10^{-2}$. The central peak with half-width $\Delta\varepsilon \sim \gamma/2$ arises due to circular dichroism, the side peaks at frequencies $\varepsilon \approx \pm\gamma_D/2$ arise due to the different circular dispersion of the medium.

b) uniform broadening prevails over Doppler $\gamma \gg \gamma_D$. For resonance detunings comparable with the homogeneous width $\varepsilon \sim \gamma$, we have

$$P_{wy}(z) = \frac{P_{wx}(0)}{4} \frac{G_+^2}{4} \frac{\beta^2}{\varepsilon^2 + \beta^2} e^{-\kappa_0 u z} \left(1 - e^{-\kappa_0 u z}\right)^2 \qquad (7)$$

The spectral dependence of the change in the polarization of the wave is determined by the Lorentzian profile of the line; the change in polarization is due to circular dichroism;

c) the line shape is arbitrary and there are large resonance detunings $\varepsilon \gg \gamma, \gamma_D$

$$P_{wy}(z) = \frac{P_{wx}(0)}{4} \frac{G_+^2(0)}{64} \frac{\gamma^4 \gamma_D \kappa_0 z^2}{4\pi\varepsilon^6} \tag{8}$$

Far from resonance, only a rotation of the plane of polarization occurs (without deformation). This rotation is due to purely polarization changes in the refractive index.

Let's summarize, that the change in the polarization of the probe signal is directly related to the density of atoms, in resonance with laser radiation. In principle, this makes it possible to detect the density of one or another gas by the value of the angle of rotation. Since the rotation of the plane of polarization takes place only in the presence of an intense pulse, the local diagnostics of the density of atoms can be carried out in this way.

It is clear from the above, that the mechanism of light polarization is directly related to the shape of the spectral line of atoms. Below are presented various mechanisms of broadening of the spectral lines of neutral atoms in a plasma.

The spectral lines of neutral atoms in plasma are broadened due to 4 main reasons.

1. Interaction of atoms with the self-radiation field, which leads to radiative damping of the oscillator and to the natural linewidth ($\gamma_{nat} \sim 10^{-4} cm^{-1}$)

2. Thermal motion of atoms, which leads to Doppler broadening ($\gamma_{Dop}^{H_2} \sim 14 cm^{-1}$, $\gamma_{Dop}^{Cs} \sim 1,2 cm^{-1}$)

3. Interaction of an atom with surrounding particles (electrons, ions, atoms), which leads to line broadening due to pressure effects. ($\gamma_{hom}^{H_2} \sim 10^{-1} cm^{-1}$, $\gamma_{hom}^{Cs} \sim 10^{-2} cm^{-1}$)

4. Interactions of atoms with the magnetic field of the tokamak, which leads to Zeeman line broadening ($\gamma_{zeem}^{H_2,Cs} \sim 1,4 cm^{-1}$)

The above widths were obtained with the following data regarding the installation of the tokamak:

The torroidal magnetic field strength   H = 30 KOe

Temperature of electrons and ions T = 100 eV = 1 000 000 K

Density of electrons and ions    $N_e \sim N_i \sim 10^{13} \div 10^{14} cm^{-3}$

Density of neutral atoms    $N_{at} \sim 10^{13} \div 10^{14} cm^{-3}$

### 3. Local diagnostics of the density of neutral atoms

Let us consider, in more detail, the possibilities of local diagnostics of the density of neutral cesium atoms by the method of stimulated rotation of polarization planes. As already noted, in the absence of external magnetic fields in a plasma, it does not change the polarization of the probe linearly polarized light signal * .

*The change in the polarization of the signal due to the Faraday rotation of the plane of polarization on the electrons of the plasma becomes significant in a very dense plasma $N_e \sim 10^{19} at/cm^{-3}$ .

If the plasma is illuminated simultaneously by a probing linearly polarized signal and an intense circularly polarized pulse, then a change in the polarization of the probing signal gives information about the density of atoms (in the region of overlap of two signals). By changing the area of overlapping of weak and strong signals, we can thereby carry out local diagnostics of atoms in space. Let us evaluate the possibilities of this research method.

In a cesium plasma at temperatures of the order of 1–10 ev, the Doppler half-width reaches values of the order of $(6 \cdot 10^{-2} - 10^{-1}) cm^{-1}$, a uniform collisional width of the order of $10^{-1} cm^{-1}$. This means that with resonance detuning $\varepsilon > 10^{-1} cm^{-1}$, the change in the polarization of the probe signal occurs due to the gyrotropy of the medium, i.e., the polarization plane rotates (without its deformation) The angle of rotation of the plane of polarization can be estimated from the formula

$$\varphi = q\xi z \qquad (9)$$

where $q = \dfrac{\pi N \omega |d|^2}{6c\hbar\varepsilon}$, $\xi = \dfrac{2|d|^2}{3\hbar^2 \varepsilon^2}|E_+|^2$ is a nonlinear intensity parameter, z is the length of the resonance medium.

For estimates it is convenient to use the following expressions

$$q = 10^{-14} \frac{N}{\Delta \upsilon} (cm^{-1})$$
$$\xi = 4 \cdot 10^{-7} \frac{P(W/cm^2)}{(\Delta \upsilon)^2} (cm^{-1}) \qquad (10)$$

where $\Delta \upsilon$ is the resonance detuning in reciprocal centimeters, P is the power density of the intense wave in units of $W/cm^2$, N is the density of cesium atoms.

At detunings $\Delta \upsilon \Box 10^{-1} cm^{-1}$ and power density $P \Box 1 mW/cm^2$, the nonlinear intensity parameter $\xi$ reaches saturation ($\xi \Box 1$) and the angle of rotation per unit length is

$$\varphi = 10^{-13} N \qquad (11)$$

Under experimental conditions, without special training, it is possible to register rotation angles $\varphi = 10^{-2} rad$, i.e. the minimum detectable density of cesium atoms is

$$N_{min} \sim 10^{11} at/cm^3 \qquad (12)$$

Thus, summarizing, we can say that with local diagnostics (distance 1 cm), we can register the minimum density of cesium atoms $N_{min} \sim 10^{11} at/cm^3$. Naturally, with a change in the length of the interaction between small and strong pulses, the minimum detectable density of atoms increases inversely with the length z.

## 4. Local diagnostics of the magnetic field

Local diagnostics of the magnetic field can be carried out by combining the effects of magnetic rotation of the polarization plane (Faraday, Cotton-Mouton effects) with the effect of stimulated rotation of the polarization plane.

### a. Faraday effect

Suppose that two waves propagate in the medium along the direction of the magnetic field ( z axis ) in the opposite direction: a probing signal polarized linearly and an intense signal polarized circularly (for example, the left circle $E_+$ ). The polarization of the intense wave does not change. The transmission equations for the electric vector of the probing radiation will have the form

$$\frac{dE_+}{dz} = 2iqE_+ \frac{1+\sqrt{1+\xi}}{2\sqrt{1+\xi}} \frac{\varepsilon}{\varepsilon - \frac{4}{3}\mu_0 H - \lambda_+}$$

$$\frac{dE_-}{dz} = 2iqE_- \frac{\varepsilon}{\varepsilon + \frac{4}{3}\mu_0 H} E_-$$
(13)

where $\lambda_+ = \frac{\varepsilon_+}{2}\left(1-\sqrt{1+\xi_+}\right)$ is the Stark level shift, $\xi_+ = \frac{2|d|^2 |E_+|^2}{3\hbar^2 \varepsilon_+^2}$, $\varepsilon_+ = \varepsilon - \frac{4}{3}\mu_0 H$. It can be seen from the transmission equations that if the wave is polarized at the input along the axis $\varphi$, then at the output its polarization plane will rotate through an angle equal to

$$\varphi = qz\left[\frac{\varepsilon}{\varepsilon - \frac{4}{3}\mu_0 H \sqrt{1+\xi}} - \frac{\varepsilon}{\varepsilon + \frac{4}{3}\mu_0 H}\right]$$
(14)

For small nonlinearities ( $\xi_+ < 1$ ) and large detunings ( $\frac{\mu_0 H}{\varepsilon} \ll 1$ ), the rotation angle $\varphi$ is equal to

$$\varphi = qz\left[\frac{4}{3}\frac{\mu_0 H}{\varepsilon} + \xi_0 + 4\xi_0 \frac{\mu_0 H}{\varepsilon}\right]$$
(15)

where $\xi_0 = \frac{2|d|^2 |E_+|^2}{3\hbar^2 \varepsilon^2}$

It can be seen from the obtained formula that the rotation in the medium has a complex structure. The first term in formula (15) describes the rotation only from the magnetic field, and the second term - only from the intense wave, and the third term - from the magnetic field and the intense wave simultaneously. For a given intensity of a powerful pulse and the strength of the magnetic field, the frequency of the weak probing field can be chosen in such a way that these terms in the rotation of the polarization plane give a contribution of the same order per unit length. Then, by crossing intense and probing waves, one can obtain the value of the gas density and the magnetic field strength at the crossing point. For example, for the values of the dimensionless parameter of the intensity of a powerful pulse $\xi_0 = 0.5$ of the magnetic field strength $H = 3KOe$ ($\Delta \nu_{zeem} = 0.14 cm^{-1}$), the frequency of the probing field should be chosen on the order of $0.1 \div 0.2 cm^{-1}$.

### b. Cotton – Mouton Effect

Suppose, that in the medium in the direction perpendicular to the direction of the magnetic field, two waves propagate in opposite directions - the probing and intense waves probing - linearly at an angle of 45degree to the z axis. The polarization of an intense wave does not change during its passage

The reduced Maxwell equations for the linear components y and y of the probing signal will have the form

$$\frac{dE_z}{dy} = 2iqE_z \frac{1+\sqrt{1+\xi_1}}{2\sqrt{1+\xi_1}} \frac{1}{\varepsilon_0 - \frac{2}{3}\mu_0 H - \lambda_1} + \frac{1+\sqrt{1+\xi_2}}{2\sqrt{1+\xi_2}} \frac{1}{\varepsilon_0 + \frac{2}{3}\mu_0 H - \lambda_2}$$

$$\frac{dE_x}{dy} = iqE_x \frac{1+\sqrt{1+\xi_1}}{2\sqrt{1+\xi_1}} \frac{1}{\varepsilon_0 - \frac{4}{3}\mu_0 H - \lambda_1} + \frac{1+\sqrt{1+\xi_2}}{2\sqrt{1+\xi_2}} \frac{1}{\varepsilon_0 + \frac{4}{3}\mu_0 H - \lambda_2} \quad (16)$$

$$\lambda_{1,2} = \frac{\varepsilon_{1,2}}{2}\left(1 - \sqrt{1+\xi_{1,2}}\right), \quad \xi_{1,2} = \frac{2|d|^2 |E_z|^2}{3\hbar^2 \varepsilon_{1,2}^2}, \quad \varepsilon_{1,2} = \varepsilon \mp \frac{2}{3}\mu_0 H$$

From the transmission equations (16) it is easy to find that the angle of rotation of the plane of polarization is equal to

$$\varphi = q \frac{1+\sqrt{1+\xi_1}}{2\sqrt{1+\xi_1}} z \left[\frac{1}{\varepsilon_0 - \frac{2}{3}\mu_0 H - \lambda_1} - \frac{1}{\varepsilon_0 - \frac{4}{3}\mu_0 H - \lambda_1}\right] +$$

$$\frac{1+\sqrt{1+\xi_2}}{2\sqrt{1+\xi_2}} z \left[\frac{1}{\varepsilon_0 + \frac{2}{3}\mu_0 H - \lambda_2} - \frac{1}{\varepsilon_0 + \frac{4}{3}\mu_0 H - \lambda_2}\right] \quad (17)$$

For small nonlinearities ($\xi_{1,2} < 1$) and a small parameter $\frac{\mu_0 H}{\varepsilon} \ll 1$, we have

$$\varphi = qz\left[\frac{8}{3}\left(\frac{\mu_0 H}{\varepsilon}\right)^2 - \frac{20}{9}\left(\frac{\mu_0 H}{\varepsilon}\right)^2 \xi\right] \quad (18)$$

where $\xi = \frac{2|d|^2 |E_z|^2}{3\hbar^2 \varepsilon^2}$.

The situation for diagnostics in this case is similar to the previous case, with the only difference, that the rotation of the polarization plane of the probe signal due to the magnetic field is due to the Cotton-Mouton effect, which is weaker than the Faraday effect

At temperatures of 1 - 10 eV, point density diagnostics can be carried out at gas densities of $10^{12} - 10^{13} cm^{-3}$ (rotation angle of the order of $10^{-2}$ rad) and magnetic field values of several kilooersteds

## Conclusions

It can be seen from the obtained results, that the rotation in the medium has a complex structure. The first term in formula (15) describes the rotation only from the magnetic field, the second term - only from the intense wave, and the third term - from the magnetic field and the intense wave simultaneously. For a given intensity of a powerful pulse and the strength of the magnetic field, the frequency of the weak probing field can be chosen in such a way, that these terms in the rotation of the polarization plane give a contribution of the same order per unit length. Then, by crossing intense and probing waves, one can obtain the value of the gas density and the magnetic field strength at the crossing point.

## Acknowledgements

KBO thanks NSP of the Slovak Republic for support